%% file: MorphoAlignment.tex
\documentclass[aps,pre,reprint,groupedaddress]{revtex4-2}

\usepackage{amsmath,amssymb,amsfonts,latexsym}
\usepackage{dsfont}
\usepackage{bm} 
\usepackage[mathcal]{euscript}
\usepackage{array}
\usepackage{lipsum}
\usepackage{wasysym}
\usepackage{multirow}
\usepackage{graphicx}
\usepackage{epsfig}
\usepackage{color}
\usepackage{xcolor}
\usepackage{ifthen}
\usepackage{tcolorbox}
\usepackage[activate=normal]{pdfcprot}  
\usepackage[colorlinks=true, linkcolor=blue, citecolor=blue]{hyperref}
\include{Booleans}
\include{NewCommands}

\usepackage{lineno}
\begin{document}
\graphicspath{{./Figures/}}

\makeatletter
\setlength{\@fptop}{0pt}
\makeatother

\title{Aggregating swarms through morphology handling design contingencies:\\from the sweet spot to a rich expressivity}

\author{Jeremy Fersula}
\affiliation{\isir}

\author{Nicolas Bredeche}
\affiliation{\isir}
\affiliation{\su}

\author{Olivier Dauchot}
\affiliation{\gul}

\date{\today}

\begin{abstract}
Morphological computing, the use of the physical design of a robot to ease the realization of a given task has been proven to be a relevant concept in the context of swarm robotics. Here we demonstrate both experimentally and numerically, that the success of such a strategy may heavily rely on the type of policy adopted by the robots, as well as on the details of the physical design. To do so, we consider a swarm of robots, composed of Kilobots embedded in an exoskeleton, the design of which controls the propensity of the robots to align or anti-align with the direction of the external force they experience. We find experimentally that the contrast that was observed between the two morphologies in the success rate of a simple phototactic task, where the robots were programmed to stop when entering a light region, becomes dramatic, if the robots are not allowed to stop, and can only slow down. Building on a faithful physical model of the self-aligning dynamics of the robots, we perform numerical simulations and demonstrate on one hand that a precise tuning of the self-aligning strength around a sweet spot is required to achieve an efficient phototactic behavior, on the other hand that exploring a range of self-alignment strength allows for a rich expressivity of collective behaviors.

\end{abstract}

\maketitle


\section{Introduction}
\label{sec:intro}

Many different aspects contribute to the design of a robot : its morphology (shape, structure, materials), sensory apparatus, motor system, control architecture, etc.~\cite{siciliano2008springer,hamann2018swarm}. All of these aspects interact and jointly determine the robot’s behavior. By designing the morphology of the robot to obey certain physical properties, one can alleviate the need for computation to achieve a desired behavior as the laws of physics "work" in favor of succeeding the task~\cite{muller2017morphological}. This design principle of using morphology to facilitate control has been proven effective in achieving complex tasks for many different robotic systems~\cite{wang2022control, shah2022tensegrity}, including swarm robotic ones~\cite{li2019particle, gros_segregation_2009, hao_controlling_2023, ben_zion_morphological_2023}.

As this design principle heavily relies on physics, a great source of inspiration lies in the field of active matter - a field for which the systems of interest look more and more alike to swarm robotic systems~\cite{ning_macroscopic_2024}. In both cases, one takes interest in the collective behaviors of large assemblies of mobile units, with a focus on control and tasks realization in the case of swarm robotics~\cite{csahin2004swarm, brambilla_swarm_2013, yang_grand_2018,  nedjah_review_2019, dorigo_swarm_2021} and a focus on the spontaneous emergence of symmetry broken phases in the context of active matter~\cite{ramaswamy2010mechanics, cates_motility-induced_2015,gompper2020,baconnier2025self}. 
While in both cases, collective behaviors emerge from local interactions there are important paradigm differences. In swarm robotics the mobile units are robots usually equipped with sensors and the ability to perform numerical computation to make decisions about their future behavior. Designing proper policies, and collective strategies to control the swarm is a central goal of swarm robotics. In the field of active matter, although many systems of interest are made of living agents performing similar decision making and cognitive processes, the interactions among the agents are described in an effective way at the level of the physical degrees of freedom, and add up to standard physical interactions such as collisions. The latter are typically avoided in the context of swarm robotics~\cite{brambilla_swarm_2013,tarapore2020sparse}. 

\begin{figure*}[t!]
    \centering
    \includegraphics[width=0.95\linewidth]{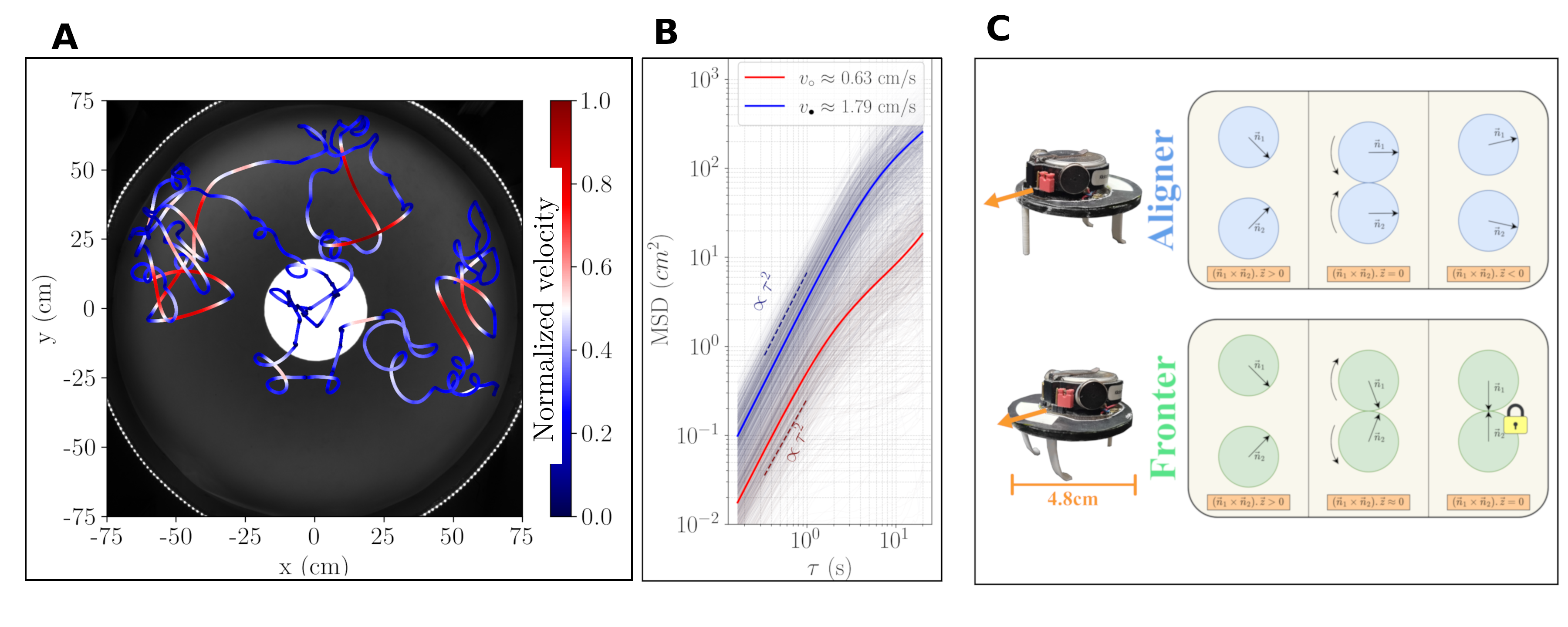}
    \caption{Experimental realization of a phototactic aggregation task for $N=64$ robots with morphological actuation through a force re-orientation response.  \textbf{A.} Each robot is programmed to perform a Run-And-Tumble motion, with different speeds based on the local light intensity. The run phases in the dark are clearly visible with a greater speed and persistence length. After some time at the boundary, robots eventually get re-injected in the bulk due to the orientational noise. Note that the typical time spent at the boundary is slightly greater for fronters than aligners due to the morphological response ($\approx$ x1.4). \textbf{B.} Effective velocity of robots (aligners) in the dark and light region, measured from the mean square displacement. \textbf{C.} Kilobot \cite{rubenstein_kilobot_2012} Robots are embedded in exoskeletons replacing their base legs. The contact point asymmetry and uneven mass distribution lead to a re-orientation response when subjected to external forces. The aligner aligns towards external forces while the fronter aligns in the opposite direction. During a collision, morphological actuation plays a role for robot aggregation : the force re-orientation response of the fronter favors aggregation, provided the intensity of said response is low enough to not cause an unending interlock.}
    \label{fig:fig1}
\end{figure*}

In the recent years, there has nevertheless been a convergence of interest, with not only more works in active matter dealing with smart agents~\cite{sugawara1997cooperative,sugawara2006collective, li2021programming, nava2018markovian, savoie2019robot}, but also works in swarm robotics demonstrating for instance that swarm aggregation benefits from collisions between robots \cite{gros_segregation_2009, hao_controlling_2023, ben_zion_morphological_2023}. More specifically it was shown in~\cite{ben_zion_morphological_2023} that a specific re-orientation response to external forces of the robots, the so-called self-alignment~\cite{baconnier2025self}, can improve or degrade aggregation for a phototactic task. In this task, robots explore their environment to find a light region the size of which is too small to fit the entire swarm, causing frequent collisions and re-orientations. In this set of experiments the robots controller allows for a complete stop. This has two consequences : (i) each robot can individually achieve the task, (ii) when stopping at the boundary of the light region the robots build up a wall preventing the remaining robots to access the light region. The success of the aggregation in the light region precisely depends in the way these remaining robots morphologically respond to this repulsive barrier.

Here we address a more subtle question, namely the one of a possible collective effect of the morphological design of the robots in achieving the task. To do so we impose the robots to maintain a finite velocity preventing a single robot alone to achieve phototaxis. Experimentally, we consider two populations of robots with opposite self-alignment: the so-called aligners align on the force they experience; the so-called fronters anti-align in the opposite direction. We observe that the contrast in task achievement is even more dramatic : while the aligners achieve no phototaxis at all, the fronters leverage a collective effect called motility-induced phase separation (MIPS)~\cite{cates_motility-induced_2015} in the realm of active matter, to successfully aggregate in the light. We further investigate the role of self-alignment for aggregating in the light using numerical simulations. By tuning continuously the self-alignment, we show that minute change in the design actually drastically alter the performance in achieving phototaxis, as one would expect in light of recent results~\cite{musacchio2025self}, but also that exploring a wider range of self-alignment strength provides an opportunity for designing highly expressive swarms in terms of accessible collective behavior.

\section{Results}
\label{sec:wall}

\subsection{Real robots experiments}
The experimental set up has been described elsewhere~\cite{ben_zion_morphological_2023} and we only recast here its main features.  A swarm of $N=64$ Kilobots~\cite{rubenstein_kilobot_2012}, 
enhanced by a 3D-printed exoskeleton replacing their native legs, evolve in a circular arena of diameter ($\diameter = 1.5m$) to perform a simple phototactic task: aggregating in a light region located at the center of the arena and covering $\sigma=6\%$ of the total area. The boundary between the light and dark regions is sharp and no light gradients can guide the robots. The light region is made purposely too small to fit the entirety of the swarm ($N_{max} \approx50$) in order to favor collisions. 

Each robot continuously monitors the local light intensity. When in the dark area, defined as an intensity below a predefined threshold, the robot executes a run-and-tumble motion with a constant velocity $v_{\bullet}$. Upon entering the lit region, where the intensity is above threshold, the robot switches to a stop-and-run type of motion, producing an effective velocity $v_{\circ}$ (see METHODS). A typical trajectory of a single robot is illustrated on~\hyperref[fig:fig1]{Fig~1.A}. The effective velocity in the dark and light regions are extracted from the tracking of the trajectories and the computation of the mean square displacement $MSD(\tau) = \langle \left(\textbf{r}(t+\tau) - \textbf{r}(t)\right)^2 \rangle $, where the brackets denote an average over the running time $t$, inside and outside the light region (\hyperref[fig:fig1]{Fig~1.B}). Averaging over many realizations of trajectories, we find a mean effective velocity ratio of one third : $v_\circ / v_\bullet \approx 1/3$.

\begin{figure*}[t!]
    \centering
    \includegraphics[width=1.0\linewidth]{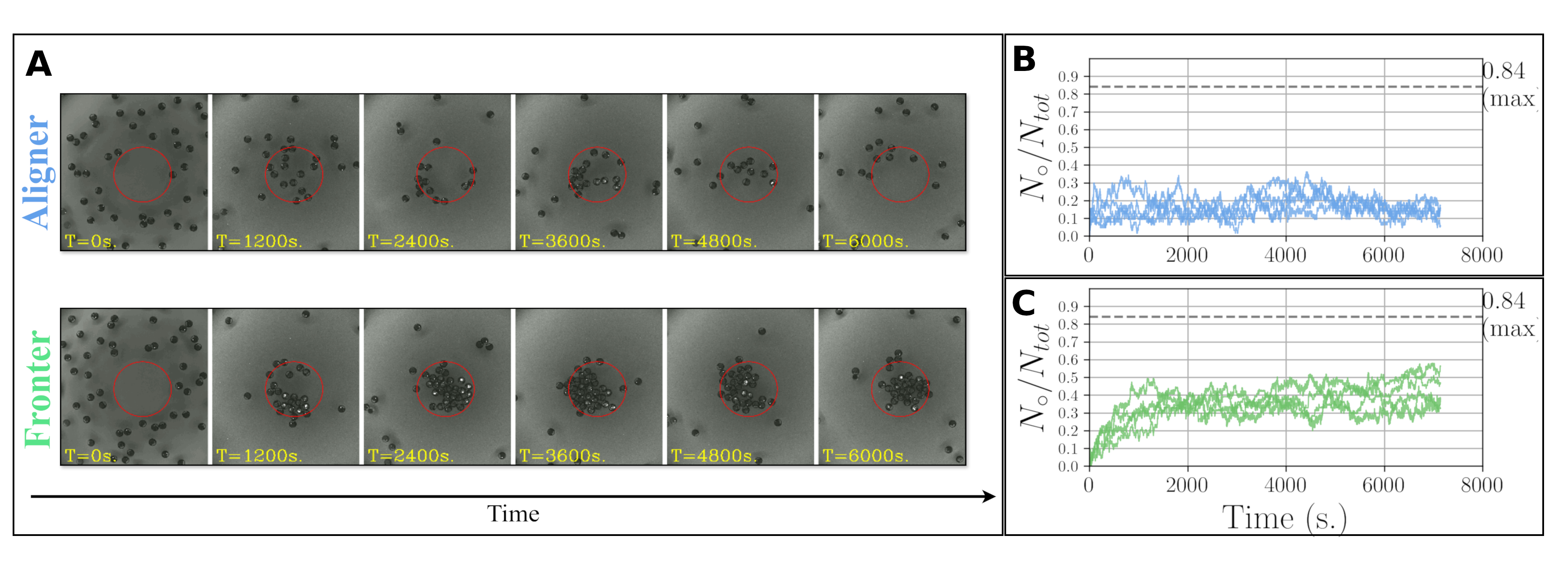}
    \caption{Experimental results for a phototactic aggregation task of $N=64$ aligners and fronters. \textbf{A.} Snapshot of real robot experiment over time for the two morphologies : aligners on top and fronters on the bottom. The illuminated area (not visible on the images) is surrounded by a red circle. The positive self-alignment of the aligners prevent them from aggregating in the light region, while the fronters succeed in clustering in and around the illuminated region. \textbf{B.} Fraction of the swarm in the light region as a function of time for 4 independent experiments quantifies the lack of aggregation for the aligners, with no significant evolution in time and a mean $N_\circ / N_{tot}$ around $0.16$. \textbf{C.} Conversely, the fronters manage to hold a consistent filling of the light region following and exponential-like convergence to a plateau value around $N_\circ / N_{tot} =0.4$.}  
    \label{fig:fig2}
\end{figure*}

Two types of exoskeleton have been designed in order to obtain two opposite self-aligning behavior (\hyperref[fig:fig1]{Fig~1.C}). The asymmetry in the legs positions and mass distribution leads to an asymmetric drag on the contact points with the floor, which in turn induces a re-orienting torque on the robot, when it experiences a force that is not aligned with its propelling direction~\cite{baconnier2025self}. For the so-called aligner morphology, two flexible legs are placed symmetrically on the back of the robot and a single rigid leg is placed centered at its front, with the mass being distributed toward the front. As a result the aligner re-orients in the direction of the force it experiences. For the fronter morphology, the two flexible legs are placed symmetrically on the front of the robot and the single rigid leg on its rear, with the mass being distributed toward the back. The fronter re-orients against the direction of the force it experiences. 

When two robots enter a "collision event", they typically experience repeated tapping~\cite{caprini_emergent_2024} and repeated re-orientation. At an effective level, the collision can be described as follow (\hyperref[fig:fig1]{Fig~1.D}): the two robots feel a force pointing outwards from the collision point. As two aligners collide, they re-orient partially away from the contact point, more-so according to the strength of morphological effects. This produce a tendency to share a common direction after a collision event. On the contrary, as two fronters collide, they re-orient partially toward the contact point, that may in turn trigger another collision. Provided that the strength of morphological effects is high enough, this cycle becomes hard to break and two colliding robots can become stuck until an external event or vibrational noise eventually lead them to escape the collision. Altogether the generic behavior of aligners, respectively fronters, is to disperse, respectively stick together, by the combined effect of activity and morphology.

At the beginning of each experiment, robots are placed randomly in the dark region of the arena with varying initial positions and orientations. A time series of image snapshots taken from a typical experiment is shown on \hyperref[fig:fig2]{Fig~2.A}, for respectively the aligners (top) and the fronters (bottom). As robots randomly explore the space of the arena, the local density quickly increases in the light region as compared to the dark one, simply because of the difference of effective velocity in the two regions. After this brief transient, one observes two clearly distinct behaviors. In the case of the fronters the initial increase of density rapidly triggers the formation of a cluster of up to 25 robots. This cluster is subject to evaporation at the boundary, with robots leaving due to orientational or positional noise driving them away from the cluster, but this evaporation is compensated by new robots randomly colliding with the existing cluster, triggering the force re-orientation response which drives the incoming robot towards the crowd.
In the case of the aligners nothing of the sort happens. Repeating 4 independent runs for each morphology and monitoring the fraction of robots inside the light region as a function of time confirm the above observation (\hyperref[fig:fig2]{Fig~2.B} and \hyperref[fig:fig2]{Fig~2.C}). In the case of the aligners, the data show no clear evolution in time, but a distribution centered around $N_\circ / N_{tot} = 0.16$, a value close to the expected proportion obtained as if there was no collisions and robots would explore the arena independently ($\sigma \frac{v_\bullet}{v\circ} = 0.18$). There is thus no sign of aggregation for the aligners. The data obtained in the case of the fronters tell a different story. One observes an exponential relaxation to a plateau taking values around $N_\circ/N_{tot} = 0.4$, highlighting the capacity of the fronters to aggregate. 

\begin{figure*}[t!]
    \centering
    \includegraphics[width=0.98\linewidth]{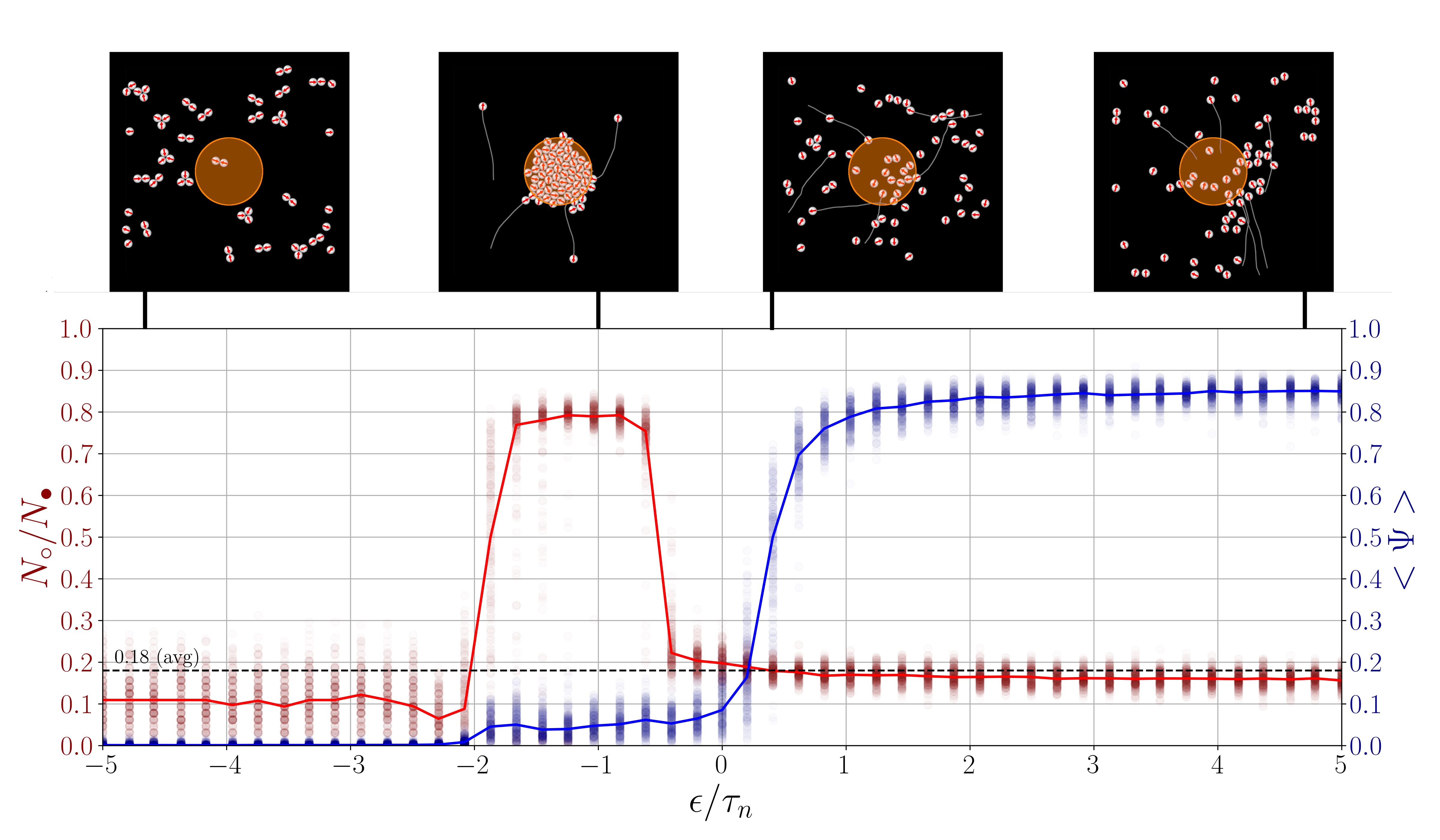}
    \caption{Fraction of the swarm in the light region (red) and total alignment (blue), averaged over the last $100\tau$ of a $1000\tau$ duration experiment, as a function of $\epsilon/ \tau_n$, for the simulated phototactic aggregation task. The horizontal dotted line is the fraction expected from a random exploration of space by non interacting agents, slowing down in the light with a velocity ratio $v_\circ/v_\bullet=1/3$. The insets display simulation snapshots of the final time state of the system at four different values of $\epsilon/\tau_n$, with typical trajectories over the last 10$\tau$.}
    \label{fig:fig3}
\end{figure*}

This effect can only be attributed to the role of morphology: aggregation is made possible through the force re-orientation response granted by the exoskeleton design. When two fronters collide and face each other, their real speed decreases to almost zero, even if their respective motors keep on running. This dramatic decrease of the velocity with the local density is the basic mechanism at the origin of the well known motility-induced phase separation~\cite{cates_motility-induced_2015}:
density increases where velocity slows down; if velocity slows down strongly enough with density, this positive feedback leads to an instability and the associated phase separation between a low density and a high density phases. In other words, the transient increase of the density induced by the "slowing down in the light" policy, coupled to the fronters' strong tendency to decrease their effective velocity during collisions, induces the seeding of the phase separation, hence the aggregation in the light region. This cannot take place for the fronters which barely slow down during a collision.

The above result demonstrates the possible cross-fertilization between the field of active matter and that of swarm robotics and stresses the importance of the morphological design for the realization of the given task. It also shows that the mechanism by which the morphology helps in the task realization may be strongly dependent on the accessible set of policies. In the case where the robots could stop in the light, the difference of success rate in the phototaxis between the fronters and the aligners results from a steric barrier taking place at the boundary between the light and the dark regions~\cite{ben_zion_morphological_2023}. In the present case, where the robots can only slow down, we just saw that it is a nucleation dynamics seeded in the light region that assists the fronters aggregation.

\subsection{Tuning self-alignment in silico}
Optimizing the morphology in view of a given task is in general a very tedious and time consuming task, because of the highly non linearity dependence of the output over many parameters. In the present case, the mechanical rules linking the morphology (friction and mass distribution) to the self-alignment strength is known(see~\cite{baconnier2025self}-appendix), and we can concentrate on the numerical evaluation of the right amount of self-alignment needed to optimize aggregation.  

The dynamics of self propelled agents, such as the robots, taking their momentum from an external~\cite{deseigne_collective_2010,deseigne_vibrated_2012} or an internal vibration~\cite{baconnier2022selective} source and including self-alignment~\cite{baconnier2025self}, is well described by a set of stochastic equations for the center of mass $\vec{r}$ and for the body orientation encoded in a unit vector $\vec{n}$~\cite{lam_self-propelled_2015}. With the exoskeleton used in the present study, one can safely assumes overdamped dynamics~\cite{fersula_self-aligning_2024}, and these equations read for each robot: 
\begin{align}
    \frac{d\vec{r}}{dt} &= v_a\vec{n} + \vec{F}_{ext} \label{equation1}\\
    \tau_n\frac{d\vec{n}}{dt} &= \epsilon(\vec{n} \times \vec{v}) \times \vec{n} + \sqrt{2D}\xi \vec{n}_\perp
    \label{equation2}
\end{align}
where $\vec{F}_{ext}$ describes the collisions with the other robots and $\xi(t)$ is a delta-correlated Gaussian noise. The collisions with the other robots are here implemented via the use of a repulsive hard core Weeks-Chandler-Andersen potential~\cite{andersen1972roles}. Time and length have been rescaled using the robot diameter $d$ as unit length and $\tau=d/v_a$ as the unit time, where $v_a$ is the typical active velocity, chosen here as the velocity in the dark $v_\bullet$. As in the real robot experiment, we perform a phototactic aggregation task keeping the same ratio $\sigma = 6\%$ for the area of the light region to the arena size, which here has periodic boundary conditions. The ratio $v_\circ / v_\bullet$ is chosen to mimic the experimental value of $1/3$, we set D=0.01 and we concentrate on varying $\epsilon/\tau_n$, where $\epsilon$ sets the sign of the self-alignment, $\epsilon=+1$ for aligners and $\epsilon=-1$ for fronters, and $1/\tau_n$ sets the strength of the self-aligning torque. In the limit where $\tau_n \rightarrow +\infty$, the self-aligning torque vanishes and the dynamics is that of standard Active Brownian Particles~\cite{romanczuk_active_2012}.

We run simulations lasting $1000\tau$, uniformly varying the parameter $\epsilon/\tau_n$ in the range $[-5, 5]$ and record the fraction of the swarm in the light region averaged over the the last 100$\tau$, as shown in red on Fig.~\hyperref[fig:fig3]{Fig.3}. The most salient feature of the result is the presence of a sweet spot of values for the parameter $\epsilon/\tau_n$ for which aggregation in the light takes place. 
In this regime, the slowing down of the dynamics inside the light region leads to a density increase, which triggers the formation of a cluster in a motility induced phase separation like scenario:
once a cluster seed is large enough, the probability of wandering robots to come in contact and aggregate with it exceeds that of a robot at the boundary to leave the cluster, therefore maintaining the growth of the aggregate. This is the same effect as the one observed in our real-robots experiments. And indeed, this sweet spots takes place for negative values of $\epsilon/\tau_n$, that is for a fronter behavior. Also in line with our observations, we find that aligners can never achieve aggregation and the fraction of agents in the light matches that of a random exploration of space, with a the ratio of velocity $v_\circ / v_\bullet = 1/3$. 

Even more surprising is the fact that increasing the magnitude of the negative self-alignment rapidly turns counterproductive. As a matter of fact, we even observe reverse phototaxis: the fraction of agents inside the light is smaller than expected from the random exploration of space. Similarly increasing $\epsilon/\tau_n > 0$ does not improve and even slowly deteriorate the performance. 
The reasons for these features can be inferred by a closer look at the late-stage state of the system (see insets of \hyperref[fig:fig3]{Fig.3}), and match with the conclusions of a recent numerical work~\cite{musacchio2025self}, where the stability of the motility induced phase separation with respect to self-alignment is analyzed.
For large and positive self-alignment, the collisions together with the self-alignment lead to collective, unidirectional motion~\cite{lam_self-propelled_2015}. This is further evidence by  measuring the swarm alignment over the last $100\tau$ $<\Psi> = |\sum_{i=0}^{64}\vec{n_i}|$, as shown in blue on ~\hyperref[fig:fig3]{Fig.3}. For $\epsilon/\tau_n > 0$, particles start to align and the value of $<\Psi>$ grows up to a plateau for which the alignment is maximal with respect to the orientational noise D and the total density. For large, but negative self-alignment, micro-clusters form everywhere, even in the low density dark region: the robots entering in contact quickly lock in a very stable binary or ternary configuration. These two emergent behaviors act against the aggregation in the light.

The results of these simulations are thus two-fold. First, comparing with the real robots experiments, one clearly sees that the morphologically rooted force-reorientation observed in the real robot systems is well-replicated by the self-alignment term, with the parameter $\epsilon/\tau_n$ driving entirely the large scale dynamics and the success in aggregation. Second, increasing the range of this parameter, we observe new large scale collective behaviors, ranging from local clustering to collective motion, suggesting a richer expressivity of the control parameter than initially expected. 

\section{Discussion}
This work demonstrates that morphological design does not merely modulate individual robot responses to environmental cues, but can fundamentally reshape the collective mechanisms through which a swarm achieves a task. By constraining robots to maintain a non-zero velocity, we remove the possibility of single-agent phototaxis and force the swarm to rely on collective effects. In this regime, the two morphologies that previously differed only moderately in performance exhibit radically divergent behaviors: aligners fail to accumulate in the light region, while fronters trigger a phase-separation mechanism reminiscent of motility-induced phase separation, driven by collision-induced velocity reduction. Morphology thus acts not simply as a facilitator of control, but as an enabler of alternative collective pathways to task success.
Importantly, the mechanism enabling aggregation differs qualitatively from previous observations under stop-and-run control policies where aggregation is limited by steric arching and boundary effects, and morphology modulates how remaining agents negotiate the crowd barrier. This highlights that the benefits of morphological computation depend critically on the space of available behavioral policies, suggesting that morphology cannot be optimized independently of control—and may even compensate for policy limitations.

Our numerical results further reveal that successful aggregation requires tuning self-alignment within a narrow “sweet spot”: too weak, and collisions are insufficient to seed clustering; too strong, and either velocity alignment or stable micro-clusters outside the light region hinder aggregation. This sharp sensitivity underscores the high-dimensional trade-offs inherent in morphology design: robustness may require morphologies that adapt dynamically rather than static, single-purpose exoskeletal structures.

More broadly, these results position morphology as a control parameter enabling a richer design vocabulary for swarm robotics. Exploring a continuous space of self-alignment strengths yields a spectrum of collective behaviors—from global flocking to jamming into localized cluster formation—suggesting that morphological tuning could be leveraged not merely for performance optimization, but for programmable expressivity. This resonates with recent theoretical developments in smart active matter, where physical interactions provide implicit computation and decision-making capacity at the collective level.

Finally, our findings raise several engineering and scientific perspectives.
(i) Toward adaptive or reconfigurable morphology: dynamically tuning mass distribution, friction, or contact asymmetry could allow swarms to switch between collective modes in response to context, much as biological collectives modulate adhesion or polarity. (ii) Learning morphology–policy couplings: combining morphological effects with decentralized learning or embodied evolution may enable robots to autonomously discover how to exploit physical interactions rather than treat them as constraints~\cite{fagan_self-organised_2018, bredeche_embodied_2018, paul_physical_2025}. (iii) Bridging active matter theory and swarm design: the observed behaviors provide a real-robot platform for testing predictions from kinetic theories of self-aligning active particles~\cite{levay_cluster_2025, jung_kinetic_2025}.

\section{Acknowledgments}
This work is supported by the MSR and SSR projects funded by the Agence Nationale pour la Recherche under Grant No ANR-18-CE33-0006 and ANR-24-CE33-7791. We thank M.Y. Ben Zion for the design of the exoskeletons, first presented in~\cite{ben_zion_morphological_2023}.

\section{Methods}
\subsection{Experimental setup}
\subsubsection{Kilobot robot}

The Kilobot robot is a 33mm wide robot consisting of a circular board mounted on 3 rigid metal legs and supporting a cylindrical battery, making it approximately 34mm tall.
On top of the board are placed a light intensity sensor, a RGB led, two vibration motors and the removable 3.7V lithium-ion battery. On the bottom of the circuit board, the robot is equipped with a close range infrared emitter/receiver, used for both the communication inter-Kilobot and computer-Kilobot.

\subsubsection{Aligner and Fronter Exoskeletons}
Exoskeletons are tripod based design printed using a photopolymer based 3D printer Stratasys Connex3 Objet260.
Each print comes embedded in a soft material, the support, which has to be removed mechanically. Materials used for both morphologies are "Support 706B" and "VeroClear". Once the support has been removed, exoskeletons are gently washed with water and a black and white detection pattern is glued on top.

\subsubsection{Arena}

The whole arena rests on a thick slab of chipboard wood of 5x200x200cm on which arena floor, walls and overall structure is screwed in. The floor consists of a 2mm thick black PMMA slab, and the walls made of a 0.2mm thick circular aluminum sheet of 1.5 meters in diameter. A main beam structure is fixed along a bearing wall of the experimentation room, two beams are placed on the opposite sides to hold black fabric covering the whole arena. The inside of the arena is lit via LEDs light strips attached to the inner part of the circular wall. About 2 meters higher than the floor, a black and white camera is set to film the entirety of the arena, and a small video projector allows for projection on part of the arena. The projected zones are all set to use only the green channel, as it matches better the Kilobot sensor intensity peak than the color blue. In front of the camera, we placed a red long-pass filter, cutting wavelengths shorter than 645nm. Doing so, the contrast between the lit and unlit regions disappears on the film, the overall light intensity is enough to track the robots and the real contrast in light intensity is maintained high enough to allow for phototaxis.
\subsection{Algorithms}
The algorithm used is an algorithm based on a \textit{sense-act} cycle, in which the robot senses its environment (light intensity) and act by driving its motor vibration intensity according to a predefined Run-And-Tumble behavior. After each cycle, an assessment is made to switch from a high speed Run-And-Tumble to a low speed Run-And-Tumble.
In the light integration phase (sense), the robot performs batches of light intensity measurements to average out sensor noise over the course of $\approx1s$. In the motion phase, the robot decides its velocity for the Run-And-Tumble. A high speed, Run-And-Tumble is performed by setting both motors of the Kilobot to $130 / 255$ for a fixed duration of $6s$, and tumbling by applying a less powerful vibration intensity to one motor only drawn at random ($110/255$). The tumbling duration is drawn from a long-tail distribution with 6 possibilities, from $1s$ to $7s$. The low speed Run-And-Tumble on the other hand cannot be implemented by reducing motor intensity only, as motor have a starting threshold below which they do not vibrate. Instead, the motion is interrupted quickly, following phases of Run-And-Tumble and complete stop with a 1 to 15 duration. At the end of the sense-act cycle, a flag is updated according to the measured light intensity to switch from one behavior to the other.
\subsection{Numerical simulations}

\label{methods:simu}

Simulations are ran using LAMMPS  \cite{thompson_lammps_2022}, an open-source software designed for molecular dynamics simulation. This software has been extended with our equations of motion, using code available publicly on Github (\url{https://github.com/jeremyaqp/AggregatingSwarms2026}). Collisions are soft and computed with a truncated and shifted Lennard-Jones potential (WCA) cut at 1.122. Initially, particles are placed randomly in the simulation a box. A first phase of Brownian motion with increasing potential cutoff from 0 to the final value is performed to avoid inter-penetration of particles. Only then the true simulation is started. Exact parameters for the simulations are as follow :

\begin{itemize}
  \setlength\itemsep{-0.4em}
    \item N=64
    \item Square L=27.7 pbc, lit region is a circle $\diameter=3.83$ ($\sigma \approx 6\%$)
    \item dt = 0.001, 1000$\tau$
    \item $\tau_v=0.001$, D=0.01
    \item $v_\circ = \frac{1}{3}v_\bullet$
\end{itemize}

The data is analyzed through python scripts available on the same Github repository. Simulation snapshots are made using Ovito \cite{ovito}.

\bibliography{Biblio/Alignment,Biblio/Collisions,Biblio/Generic-ActiveMatter,Biblio/Generic-SwarmRobotics,Biblio/Other-Conclusion,Biblio/Other-Introduction,Biblio/SwarmAggregation,Biblio/Tools,Biblio/misc.bib}

\WIPbox{ 
\todo{\textbackslash todo\{text\}}\\
\warn{\textbackslash warn\{text\}}\\
\dont{\textbackslash dont\{text\}}\\
\note{\textbackslash note\{text\}} \\
 \jf{\textbackslash jf\{text\}} - \nb{\textbackslash nb\{text\}} - \od{\textbackslash od\{text\}}
\begin{itemize}
    \item Mettons-nous des cases aux sous-figures ? (FIG1 vs FIG2) quoique le journal ne prends peut-être que les sous-figures
    \item  Le schéma FIG1.D est-il adapté et bien construit ?
    \item  Doit-on justifier l'arrivée à un steady state pour les simulations ?
    \item Doit-on augmenter le nombre de simus ? (compter 24H pour les données de FIG3, voir methods) On verra probablement une branche vers $\epsilon / \tau_n \approx -1.5$, les cas où le cluster se forme dans le noir.
    \item  FIG3. Possibilité de superposer l'alignment $<\psi>$ pour contraster
    \item Le "small dent" des Fronters à $\epsilon / \tau_n \approx -2.5$ n'est pas adressé
\end{itemize}

}

\end{document}

%% file: Booleans.tex
\newboolean{WIP}
\setboolean{WIP}{false} 

%% file: NewCommands.tex
\newcommand{\od}[1]{\begingroup\color[rgb]{1,0,1}#1\endgroup}
\newcommand{\nb}[1]{\begingroup\color[rgb]{0.5,0,0.5}#1\endgroup}
\newcommand{\jf}[1]{\begingroup\color[rgb]{0.48,0.18,0.04}#1\endgroup}
\newcommand{\gul}{Gulliver UMR CNRS 7083, ESPCI Paris, PSL Research University, 10 rue Vauquelin, 75005 Paris, France}
\newcommand{\isir}{Institut des Systèmes Intelligents et de Robotique, Sorbonne Universit\'{e}, CNRS, ISIR, F-75005 Paris, France}

\newcommand{\su}{Laboratoire Jean Perrin, IBPS,  Sorbonne Universit\'{e}, CNRS, F-75005 Paris, France}

\newcommand{\beq}{\begin{equation}}
\newcommand{\eeq}{\end{equation}}
\newcommand{\ben}{\begin{equation*}}
\newcommand{\een}{\end{equation*}}
\newcommand{\bseq}{\begin{subequations}}
\newcommand{\eseq}{\end{subequations}}
\newcommand{\bea}{\begin{eqnarray}}
\newcommand{\eea}{\end{eqnarray}}
\newcommand{\bal}{\begin{align}}
\newcommand{\eal}{\end{align}}

\newcommand{\WIPbox}[1]{%
  \ifthenelse{\boolean{WIP}}{%
    \begin{tcolorbox}[colback=yellow!10,colframe=black,title={Work in Progress Notes}]
      #1
    \end{tcolorbox}
  }{}%
}

\ifthenelse{\boolean{WIP}}{
    \newcommand{\note}[1]{%
      \sbox0{\textbf{NOTE #1}}%
      \ifdim\wd0>0.9\linewidth
        \textbf{\colorbox{green!30}{\parbox{0.9\linewidth}{NOTE #1}}}%
      \else
        \textbf{\colorbox{green!30}{NOTE #1}}%
      \fi
    }
    \newcommand{\dont}[1]{%
      \sbox0{\textbf{DONT #1}}%
      \ifdim\wd0>0.9\linewidth
        \textbf{\colorbox{red!40}{\parbox{0.9\linewidth}{DONT #1}}}%
      \else
        \textbf{\colorbox{red!40}{DONT #1}}%
      \fi
    }
    
    \newcommand{\todo}[1]{%
      \sbox0{\textbf{TODO #1}}%
      \ifdim\wd0>0.9\linewidth
        \textbf{\colorbox{yellow!50}{\parbox{0.9\linewidth}{TODO #1}}}%
      \else
        \textbf{\colorbox{yellow!50}{TODO #1}}%
      \fi
    }
    
    \newcommand{\warn}[1]{%
      \sbox0{\textbf{WARN #1}}%
      \ifdim\wd0>0.9\linewidth
        \textbf{\colorbox{orange!40}{\parbox{0.9\linewidth}{WARN #1}}}%
      \else
        \textbf{\colorbox{orange!40}{WARN #1}}%
      \fi
    }
}{
  \newcommand{\note}[1]{}
  \newcommand{\todo}[1]{}
  \newcommand{\warn}[1]{}
  \newcommand{\dont}[1]{}
}

